# Strategic Preemption Under Shared Catastrophic Risk: The Suicide Region and the Race to Artificial General Intelligence

David Tan[a,b,*]

**ABSTRACT**

We analyze a continuous-time preemption game with shared catastrophic externalities. When the cost of catastrophe is embedded in both players' payoffs, the risk term cancels out in the equilibrium indifference condition. This creates a "suicide region" where competitive pressures force rational agents to deploy despite negative risk-adjusted net present values. We apply this framework to the race for artificial general intelligence (AGI). We show that this suicide region widens as the cost of systemic ruin grows: higher catastrophic risk does not deter the race but instead enlarges the set of conditions under which rational actors deploy despite negative social value. We characterize the resulting welfare distortion against a social planner's benchmark and demonstrate how two complementary mechanisms – private liability and prize-sharing – can close the suicide region. Private liability raises the cost of unsafe deployment while prize-sharing reduces the strategic imperative to deploy first. "Warning shots" (sub-existential disasters) will fail to deter AGI acceleration, as the winner-takes-all nature of the race remains intact.

[*] Corresponding author, email: d.tan@mq.edu.au

[a] College of Business Administration, American University of the Middle East, Egaila, KUWAIT

[b] Department of Applied Finance, Macquarie University, Sydney, AUSTRALIA

Acknowledgements

I would like to thank Alvin Moon, Lisa Abraham, Joel Predd, seminar participants at the Center for the Geopolitics of Artificial General Intelligence (RAND Corporation), Candice Tan, and Piyanuth Saithanu for their insightful and valuable feedback. I am grateful to Sam Roggeveen for editorial input on the policy version of this paper. I would also like to thank Brandon Cachapero for his excellent research assistance. All remaining errors are my own.

Current Draft: May 2026

"*Mitigating the risk of extinction from AI should be a global priority alongside other societal-scale risks such as pandemics and nuclear war.*"

- Statement on AI Risk, Centre for AI Safety (2023) Signed by: Sam Altman (CEO, OpenAI), Demis Hassabis (CEO, Google DeepMind), Dario Amodei (CEO, Anthropic), Geoffrey Hinton (Nobel Laureate), and Bill Gates (Co-founder, Microsoft).

## 1. Introduction

How do strategic races behave when the downside of failure is catastrophic and shared across all players? Standard theory states that in a race characterized by extreme volatility in outcomes and downside risk, the rational strategy for competitors is to delay investment: the value of waiting rises with the cost of a mistake, and a rational player will likely choose to delay costly irreversible investments that may fail (Dixit & Pindyck, 1994; Weeds, 2002). However, in a race characterized by preemption where the winner gains a sizeable advantage, the prospect of coming second-place compresses the value of waiting. Even so, it remains a widely held intuition that a sufficiently large catastrophic downside will still deter premature deployment – that beyond some threshold, the sheer scale of the potential disaster restrains the race.

We show that this intuition fails when the cost of failure is borne not privately by the failing player but systemically, by all players simultaneously. As such, the cost of failure is present across all players' payoffs. When deciding whether to wait or deploy, because the catastrophic risk term appears identically in every player's payoff, it cancels out in the equilibrium indifference condition, resulting in the magnitude of the potential downside disaster having no bearing on the decision to wait or execute. This results in players racing undeterred – regardless of how large the catastrophe is – and deploying preemptively as if the risk were not there.

The race to develop artificial general intelligence (AGI) is the most consequential instance of this structure. A misaligned AGI does not impose a private cost on the developer who deploys it; the resulting catastrophe is borne by all players — winner and loser alike. The AGI race is thus a preemption race with a shared catastrophic externality that cancels out in the equilibrium indifference conditions.

Existing studies on real options theory (Grenadier, 2002; Huisman & Kort, 2003) assume that the downside of a failed investment for a player is bounded by its sunk cost of the investment ($I$). However, in the AGI race, the downside of failure extends beyond $I$. First, the payoff structure is strictly winner-takes-all. Unlike conventional R&D and patent races where the first mover simply gains a market share advantage, the first to achieve AGI will secure a decisive strategic dominance, effectively liquidating the sovereignty of the second mover across military, geopolitical and economic domains. Second, the downside risk is not merely bounded by $I$; rather, it includes a non-negligible probability of unbounded systemic ruin[1] ($D$). Together, this creates a fatalistic equilibrium.

As the second mover also bears the cost of the first mover's failure – global ruin – then the value of waiting is effectively zero. For the second mover, waiting no longer provides a hedge against the first mover's mistakes. As such, the fear of preemption will dominate both parties' calculus, and the result will be an AGI arms race and its resultant race to the bottom. Evidence has shown that there is a strategic erosion of safety standards as each actor races towards AGI in order to preclude their rival from securing unfettered domination[2]. That is, there exists an "alignment tax" where investment in AI safety testing (ensuring that AI behavior is aligned with human goals) slows down one's progress. When the geopolitical cost of being the second mover is perceived as infinite, actors rationally dismantle safety guardrails to maximize speed of AGI deployment.

---

[1] A superintelligent AGI is defined as a system that possesses the capacity to significantly surpass human cognitive performance across all economically relevant domains. Leading AI safety researchers (e.g., Russell, 2019; Metz, 2023) argue that such systems present a unique "alignment problem": the technical challenge of ensuring the AGI's objectives remain compatible with human welfare as its capabilities scale. Failure to solve this alignment problem prior to deployment creates a non-negligible tail risk of outcomes ranging from total economic destabilization to existential ruin (Bostrom, 2014). In our model, $D$ represents the monetized disutility of this unbounded downside scenario.

[2] Evidence of this safety-for-speed substitution is becoming increasingly observable. In 2024, multiple high-profile safety researchers at leading AI laboratories (e.g., OpenAI's Superalignment team) resigned, citing concerns that commercial deployment timelines were taking precedence over safety protocols (The Economist, 2024b). Moreover, strategic analyses circulating in policy circles (e.g., Aschenbrenner, 2024) explicitly characterize rigorous safety testing as an "alignment tax", and that it is a friction that the US cannot afford as it jeopardizes US AGI research progress against China.

In this paper, we mathematically derive the dynamics of the AGI arms race within a continuous-time real options framework. We augment existing game theory models in the real option space for R&D (Weeds, 2002; and Huisman and Kort, 2003) by including an endogenous ruin parameter ($D$) that is correlated with deployment velocity. We uncover the "suicide region" in the AGI race: an investment space where certain parameter settings and competitive pressures force firms to invest and deploy as quickly as possible despite projected negative net present values. If the probability of total loss is identical for the second mover regardless of their action, then the incentive to wait creates no hedge value. Consequently, the only rational strategy – even in the face of existential risk – is to race for control of the decisive strategic advantage. Even if the probability of safely controlling rapid AGI deployment is low, it still exceeds the zero probability of control afforded to the second mover. Thus, nations will rationally choose the slim chance of survival by being the first mover over the certain doom of the follower.

This paper makes four contributions to current literature. First, we show that in a preemption race where the cost of catastrophe is borne by all players alike, this cost term enters every player's payoff identically and therefore cancels out in the equilibrium indifference condition. This results in the magnitude of the catastrophic risk having no bearing on a player's decision to delay or deploy. We term this the "cancellation effect". This overturns the conventional intuition – central to the logic of mutually assured destruction – that when the size of the shared catastrophe is sufficiently large or even existential, players will restrain themselves and resist deployment. We further show that the suicide region widens monotonically in $D$: higher catastrophic risk does not deter the race but enlarges the set of conditions under which rational actors deploy despite negative risk-adjusted net present value.

Second, we formally establish the social planner's benchmark for AGI deployment ($V^*_{social}$) and find that players in the competitive game will always deploy prematurely – highlighting the inefficiency in the decentralized equilibrium. The welfare gap widens with the cost of global ruin ($D_{social}$) and narrows as safety research approaches completion. Even players who internalize the cost of global ruin will still deploy socially prematurely as the true social cost of ruin is *2D* in the two-player game.

Third, we derive the boundary conditions that eliminate the suicide region entirely. Specifically, when private liability is imposed on the players' payoffs and reaches the magnitude of *2D*, the suicide region is avoided. Alternatively, there is an optimal prize-sharing percentage (via windfall clauses), $S^*$, that nullifies the winner-takes-all characteristic of the race and closes the suicide region.

Fourth, we show that private liability – if designed correctly – acts as an *ex ante* deterrent to unsafe AGI deployment, even in the case of existential risk. Under Knightian uncertainty (Knight, 1921) over the distribution of misalignment outcomes at the moment of deployment, the Leader cannot reliably discount the existential tail risk – and *ex ante* deterrence of uncertified deployment therefore reduces the likelihood of both sub-existential and existential outcomes, as both stem from the same underlying misalignment failure. The limitation of private liability lies in *ex post* enforcement when an existential event has occurred, but its deterrent effect remains intact.

Private liability and prize sharing are complementary mechanisms that effectively address the suicide region from opposite sides of the payoff structure. Private liability increases the cost of unsafe AGI deployment, while prize-sharing reduces the incentive to deploy first. AGI research verification slows the race during the development phase but introduces instability during the endgame "breakout" stage. However, this can be mitigated with a prize-sharing instrument.

Although we develop this model based on the AGI race, its findings can be generalized to all races that are characterized by a shared risk of catastrophe and a substantial prize for the winner alone, with little upside for subsequent movers. Under these conditions, the catastrophic cost is present in every player's payoff and cancels from the equilibrium indifference condition, resulting in the size of the catastrophe has no deterrence effect. Moreover, there exists a suicide region where players will race to deploy early despite the expected risk-adjusted outcome being negative as the winner-takes-all dynamic compresses the value of waiting. Examples plausibly include the development of autonomous weapons systems, weaponization of space, and advances in synthetic biology with pandemic potential. The race towards AGI is the sharpest form of this

structure – globally shared catastrophic risk and a near winner-takes-all game – making it the most consequential instance of the game.

The remainder of the paper is organized as follows: Section 2 reviews the related literature on real options and AI safety; Section 3 presents the model setup; Section 4 derives the equilibrium strategies and the Suicide Region; Section 5 analyzes mechanism design and policy implications; and Section 6 concludes.

## 2. Related Literature

### 2.1. Real Options and Strategic R&D

Our paper builds on the long tradition of work on the strategic timing of technology races. Foundational literature – Dasgupta and Stiglitz (1980), Reinganum (1981, 1982), Harris and Vickers (1985), and Fudenberg and Tirole (1985) – laid the groundwork for preemptive races where competitive pressure induces players to deploy prematurely relative to their private optimal threshold. Competitive rent dissipation drives entry timing to the point where leader and follower payoffs equalise. The option games framework (Grenadier, 2002; Weeds, 2002; Huisman and Kort, 2003) combines preemption games with real option analysis, allowing uncertainty in the underlying asset value to interact with the competitive timing of deployment. Our study extends this lineage by integrating a shared catastrophic externality in a preemptive-timing problem. We show that when the cost of the catastrophic externality is borne symmetrically by all players, it cancels out in the equilibrium indifference equation and has no impact on the strategic timing of deployment.

In standard real option theory, the presence of sufficient uncertainty will incentivize firms to delay their irreversible investments in order to avoid costly technological errors until the uncertainty is resolved (Dixit & Pindyck, 1994). However, research on R&D races indicates that the predicted strategy to delay can be altered under specific conditions. In a winner-takes-all race, Huisman and Kort (2003) and Weeds (2002) formalized the tension between the incentive for firms to preempt their rivals in securing the monopoly prize and the incentive to delay investment and avoid the sunk costs of a failed technology. Weeds (2002) finds that when there is sufficient volatility and technological risk, the incentive to delay investment can dominate the decision to preempt a rival, leading to mutual delay in the race otherwise referred to as the "sleeping patent" equilibrium[3].

Crucially, the traditional literature assumes that the loss for the leader in such races is bounded by their sunk costs (I). However, in the AGI arms race, there exists an endogenous probability of systemic catastrophe (D) that is correlated with the velocity of technological deployment. Importantly, current theory fails to price this risk appropriately as the cost of failure – in the misaligned AGI scenario – extends beyond the firm's balance sheet to the entire system.
Real options analysis has been applied to environmental investments where firms face catastrophic downside risk (Cortazar, Schwartz and Salinas, 1998), though such studies typically bound individual losses by firm-level exposure rather than the systemic ruin parameter we introduce.

However, a recent study by Abraham, Kavner and Moon (2025) models the AGI arms race as a symmetric two-player prisoner's dilemma game and finds that a stable equilibrium of cooperation (safe AGI development) can occur when the cost of AGI misalignment risk exceeds the first mover advantage. This condition shifts the game from a prisoner's dilemma to a coordination game – where mutual restraint becomes a stable Nash equilibrium. Abraham et al (2025) model the conditions prior to the race beginning – when players can choose whether to cooperate or defect. In contrast, the preemptive game of the current paper examines whether to deploy AGI while the race is already underway, where the opponent's acceleration is the default rather than a strategic choice to be deterred. Under our continuous time preemption game, increasing $D$ does not deter players from accelerating and deploying, as waiting will result in one of two potential outcomes: loss of sovereignty and strategic agency if their opponent deploys a successful AGI, or AGI misalignment and global ruin. Given this, the only rational choice for a player is to race. Our "cancellation effect" identifies a mechanism where the risk of $D$ does not halt the race once it has begun.

---

[3] Weeds (2002) refers to a useful analogy: a marathon where runners pace one another, mutually delaying the sprint (investment) until the course ahead clears (uncertainty resolves). In this state, the option value of waiting is so high that even the threat of a rival cannot induce early exercise.

### 2.2.Catastrophic Risk, Externalities, and AGI

The current paper engages with the literature on the economics of externalities and catastrophic risk. The conventional concern in public economics studies – that costs imposed by private actors on third parties remain uninternalized (Pigou, 1920; Cornes and Sandler, 1996) – takes a distinct form when the downside is catastrophic, correlated across actors, or characterized by deep uncertainty. Weitzman (2009) and Gollier, Jullien and Treich (2000) show that aggressive precaution is warranted when there is deep uncertainty over catastrophic outcomes, as standard expected-utility calculations breakdown. Bernard, Rheinberger and Treich (2017) examine catastrophe aversion under interdependent risks, and Berger, Emmerling and Tavoni (2016) characterise optimal abatement under model uncertainty about catastrophe.

Growiec and Prettner (2025) identify a related paradox at the individual level: that as existential risk increases, impatience also rises due to higher discount rates, which reduces their investment in risk-mitigating projects. Our results are complementary: the cancellation effect operates through payoff symmetry in a preemption game rather than through time preferences – it holds regardless of players' discount rates or patience levels. That is, our results are not reliant on the time preferences of the actors. Together, the cancellation effect and the impatience dynamic uncovered by Growiec and Prettner (2025) indicate that – through multiple channels – an increase in the risk of existential catastrophe fails to halt the race.

Unlike the conventional externality setting where the players fail to internalize the cost of deployment as it is borne completely by a third party, our setup features downside risk that is imposed symmetrically across all players. As the risk term appears in all players' payoffs, these terms cancel out in the equilibrium indifference condition governing the timing of deployment, resulting in the magnitude of the catastrophe having no deterrence effect on the race.

While financial economists have applied real options theory to the R&D of standard technologies, there exists a burgeoning qualitative literature on AI safety and the potential risks of an AI arms race. Armstrong, Bostrom, and Shulman (2016) argue that intense competitive pressures will inevitably force players (nation states or large AI firms) to erode safety standards in favor of speed, a phenomenon they refer to as a "race to the precipice". The core issue with forgoing caution is an increase in the likelihood of an alignment problem between the superintelligent AGI and its creators. The alignment problem (Russell, 2019; Amodei et al., 2016) posits that an AGI agent whose objective function results in subobjectives that are misaligned with human values can result in catastrophic loss of human utility.

Unlike conventional industrial accidents (e.g. Chernobyl), the downside of a misaligned AGI could potentially be unbounded and global (Bostrom, 2014). This idea stems from the "instrumental convergence" thesis: a scenario where a superintelligent agent may rationally seek to accumulate unbounded resources (energy, computing power, financial assets) in order to maximize its objective function, even if this action results in the expropriation of all human assets.

Naudé and Gries (2022), model AGI development as a irreversible investment under uncertainty, though their study focuses on optimal timing for a single representative actor. The broader economic literature on AGI has focused on productivity growth and labour displacement (Nordhaus, 2021), rather than on the strategic dynamics of competitive deployment that we analyse in the current paper.

## 3. Model Motivation

### 3.1. The Asset: Artificial General Intelligence

This section motivates the key assumptions of our model through institutional observation. Each subsection maps an observable feature of the current AGI race to a specific parameter or structural choice in the formal model that follows.

Conventional AI is typically trained for a specific task (e.g. image recognition, large language models, pattern matching) – otherwise referred to as "narrow AI". Experts[4] believe that narrow

[4] See Bostrom (2014) for a discussion on the transition from "Oracle AI" (tools) to "Sovereign AI" (agents). See also Russell (2019) regarding the impossibility of manually specifying objective functions for general-purpose systems, and Amodei et al. (2016) for the distinction between concrete safety problems in narrow systems versus alignment stability in general systems.

AI is safer and is much less likely to escape its safety guardrails given its narrower objective functions within a more defined context. AGI, in contrast, will be able to perform all human cognitive tasks at superhuman velocity (Goertzel, 2014). As such, AGI will theoretically be more prone to containment failure, as a superintelligent agent will be better able to bypasses safety guardrails designed for narrower capabilities. From a financial economics perspective, AGI differs from conventional technological shocks in three fundamental ways.

First, AGI is expected to result in a zero marginal cost of intelligence[5]. Unlike the industrial revolution where new technologies were replacements for physical labour but still depended on human cognitive abilities, AGI will substitute human cognitive labour entirely at zero marginal cost. This will result in a super-exponential growth in the value of the asset ($V$) as economic value is decoupled from human population constraints. Second, AGI will possess the ability for recursive self-improvement as AGI systems design, create and deploy superior versions of themselves. Good (1965) refers to this phenomenon as an "intelligence explosion". This implies a non-constant volatility ($\sigma$) in asset value as – when the technology matures and self-improves – the range of potential outcomes widens[6]. Finally, the second mover's payoff is approximately zero ($S \approx 0$) as AGI confers the first mover with super intelligent capabilities that will dominate and exponentially accelerate progress across all domains, from automated research to military capabilities (Bostrom, 2014).

### 3.2. The Cost of Safety ("Alignment Tax")

In mainstream R&D models, safety testing is treated as a regulatory friction – a fixed cost or temporary delay required to ensure sufficient quality for deployment. The downside of safety testing in the traditional R&D context is simply a delay in deployment, or – at worst – a loss in the investment and potential reputational costs. This contrasts with the AGI arms race dynamics,

---

[5] This economic transformation is already observable in the deployment of Large Language Models (LLMs). The production function of intelligence shifts from a variable-cost model (human wages) to a high fixed-cost, near-zero marginal cost model (silicon). While the upfront capital expenditure to train a frontier model is substantial (often exceeding $100 million), the marginal cost of generating a unit of high-quality prose or code is asymptotically approaching zero.

[6] We model volatility ($\sigma$) as state-dependent – increasing with the value of the asset ($V$), as recursive self-improvement implies that as the system's capabilities increase, the dispersion of potential future states widens, ranging from solving intractable scientific problems to catastrophic failure modes.

where safety testing plays a key role in whether the asset value ($V$) will result in a positive value or systemic ruin ($-D$). In the context of AGI, investments in safety and capability are often viewed as substitutes, with safety protocols often referred to as the "alignment tax" – a computational and temporal burden that reduces the speed of development (Aschenbrenner, 2024).

There is evidence suggesting that, in the current AGI race between the US and China, the actors are rationally choosing to minimize the alignment tax in order to strive for maximum velocity. For example, major AI research labs have recently shut down their AI safety teams[7] (Lubetkin, 2024; Aschenbrenner, 2024, The Economist, 2024b). This dynamic is fueled by an acute fear of preemption ($L > F$) which results in players bypassing their safety guardrails in order to preclude the rival from securing a strategic (and decisive) advantage (The Economist, 2024a). This observation is consistent with the fundamental tension embedded in our model: the option to wait is systematically undervalued due to the perceived existential threat of being the second mover. If the probability of systemic ruin is equal regardless of who is the first mover, then the hedging value of the option to delay is zero. Thus, the rational actor preempts to maximize strategic agency, preferring the non-zero probability of controlling the outcome over the certainty of helplessness.

### 3.3. Benevolent Preemption (The "Savior's Trap")

Economic theory typically assumes that the agents in an innovation race are primarily motivated by the economic rents of the winner ($V$). However, in the AGI race, the game is further complicated by the presence of the "benevolent preemption" effect: a dynamic where an actor believes that the safety protocols of their rivals are strictly inferior ($\pi_{rival} < \pi_{self}$), and hence, there is a greater danger of systemic catastrophe if they themselves are not the first mover. This dynamic is empirically observable. Prominent AI figures have articulated this benevolent preemptive logic[8], justifying their participation and acceleration in the AGI arms race to prevent their rivals from achieving AGI – presumably because their rivals lack the competence or sufficient safety guardrails to contain misalignment.

[7] In May 2024, OpenAI disbanded its "Superalignment" team – a group specifically tasked with ensuring AGI safety – after the resignation of its co-leads, Ilya Sutskever and Jan Leike.

[8] This "benevolent preemption" is explicitly articulated by the leaders of major AI laboratories. Dario Amodei (CEO of Anthropic) has argued that his firm must race to the frontier of capabilities specifically to "maximize leverage" over the industry's safety standards, a strategy effectively requiring them to outpace less safety-conscious rivals.

In this setup, this benevolent phenomenon accelerates the fear of preemption and compels actors to deploy sooner under a moral imperative. The agent perceives an additional “saviour premium”: The utility gained from a reduction in the probability of systemic ruin by preventing a rival from deploying AGI. This creates a “unilateralist’s curse”, coined by Bostrom et al (2016), where individual actors rush to be the first to attain AGI to ensure control of deployment – in the belief that only they themselves can provide safe stewardship – while paradoxically increasing the aggregate risk of systemic catastrophe as the race accelerates beyond the optimal safety threshold.

## 4. The Model

### 4.1. Model Setup

First, we assume a continuous-time economy with two risk-neutral sovereign agents, $i \in \{1,2\}$, competing to develop and deploy AGI. The state of the world is defined by the economic value of the AGI asset, denoted by $V_t$. In our model, $V_t$ follows a Geometric Brownian Motion (GBM) within a risk-neutral world $\mathbb{Q}$:

$$dV_t = \mu V_t dt + \sigma V_t dZ_t \tag{1}$$

where $\mu = r - \delta$ is the risk-neutral drift parameter, representing the expected growth rate in AI capabilities;

$\sigma$ is the volatility parameter, representing technological uncertainty;

$dZ_t$ is an increment in the Wiener process.

---

Similarly, Elon Musk launched xAI with the stated goal of building a “maximum truth-seeking AI” to counter the perceived misalignment of competitors like OpenAI and Google, framing his entry into the race as a necessary corrective intervention. Sam Altman (CEO of OpenAI) has frequently justified the company’s deployment strategy as a means to “iteratively deploy” and “acclimate” society, positing that a US-led democratic AGI is the only hedge against authoritarian alternatives. In all three cases, the actor justifies acceleration via a belief in superior safety competence.

As discussed in Section 3, volatility ($\sigma$) is expected to be large due to the recursive self-improvement abilities of AGI, resulting in an intelligence explosion.

Unlike conventional real option scenarios where safety testing is a fixed cost and has no impact on the underlying payoff structure; in the race for AGI, we introduce a safety learning function, $\pi(\tau)$, that links the time to deployment and the payoff of the asset. $\pi(\tau)$ is expressed as a function of time to deployment because it is assumed that safety research time ($\tau$) increases the likelihood of aligned AGI. Conversely, the likelihood of misaligned AGI – that will result in systemic ruin – is equal to $1 - \pi(\tau)$. Safety is assumed to be an increasing concave function of research time, as we assume there is diminishing returns to safety research. The probability of achieving a safe AGI that is perfectly aligned is defined as:

$$\pi(\tau) = 1 - e^{-\lambda\tau} \tag{2}$$

where $\lambda$ is the safety learning rate, assumed to be strictly larger than zero;

$\tau$ is the duration of safety research, equivalent to time to AGI deployment. This assumes waiting is directly related to learning.

(2) indicates that immediate deployment of AGI with zero safety testing ($\tau \to 0$) will almost guarantee AGI misalignment ($\pi \to 0$). If the agent waits and researches safety for an infinitely long time ($\tau \to \infty$), then the probability of AGI alignment approaches near certainty ($\pi \to 1$). This function depicts the tension in our model: agents must trade-off the prospect of being the first mover and winning the race ($V_t$) with the likelihood of causing systemic catastrophe from premature deployment.

While traditional real options analysis determines the deployment thresholds by the value-matching and smooth-pasting conditions of a Hamilton-Jacobi-Bellman equation, we model the "end-game" decision to exercise deployment by comparing the payoffs for the Leader and Follower to derive the equilibrium thresholds. As such, we use equation (1) as a tractable representation of the uncertainty of the trajectory of technological advancement, consistent with the broader real options literature, despite the diffusion process not directly driving the analytical result.

The qualitative findings of this paper would remain unchanged under alternate assumptions of the $V_t$ diffusion process – such as state-dependent volatility, jump-diffusion processes, or regime-switching models – as core results of this paper are based on the algebra of the pay-off functions at the end-game phase of the race, not the dynamic of $V_t$ itself. The GBM assumption is relevant if one seeks to derive the optimal time of AGI deployment or explicitly price the option of waiting; important questions that we reserve for future research. The diffusion process is included to situate the model within the real-options tradition and represent technological uncertainty, though the equilibrium thresholds are derived from the static strategic payoffs.

### 4.2.Payoff Structures

The race is modelled as a symmetric, continuous-time stochastic game. The game ends when a player exercises their option to deploy AGI. The time of deployment is denoted as $\tau$. The first player to deploy AGI is referred to as the Leader (*L*), and the other player who did not deploy is the Follower (*F*).

As outlined in Section 3, this game has a winner-takes-all market structure with a potential outcome of global ruin stemming from AGI alignment failure. We have the following payoff structures:

*The Leader's Payoff (L)*

The leader exercises the option to deploy AGI at time $\tau$ with a sunk investment cost of *I*. With probability $\pi(\tau)$, the AGI asset is aligned, and the leader reaps the economic benefit of $(1-S)V_t$ where $S \in [0,1]$ is the market share of the second mover (the Follower). In the strict winner-takes-all dynamics of the AGI game, we expect $S \approx 0$. On the other hand, there is a $1-\pi(\tau)$ probability of misalignment and a global systemic catastrophe with a monetized utility cost equal to *D*.

$$L(V_\tau,\tau) = (1-S)\pi(\tau)V_t - \big(1-\pi(\tau)\big)D - I \qquad (3)$$

*The Follow's Payoff (F)*

Once the Leader deploys AGI, the Follower's payoff is determined by the second-mover dynamics of the game. If AGI is aligned, the Follower holds a market share of $S$ with probability $\pi(\tau)$. However, in the event of misalignment – with probability $1 - \pi(\tau)$ – the Follower incurs a monetized disutility penalty equal to *D*, just as the Leader would. Regardless of the Follower's inaction, they too are subject to penalty *D* as AGI misalignment is a global event.

$$F(V_\tau, \tau) = S\pi(\tau)V_t - \big(1 - \pi(\tau)\big)D \quad (4)$$

The "cancellation effect" is the fact that the ruin term, $-\big(1 - \pi(\tau)\big)D$, is present in both (3) and (4). This reflects the shared fate of both players: waiting does not confer any immunity to the downside of AGI misalignment. This symmetry in the payoff structure is what drives the equilibrium results in the following section.

### 4.3. Equilibrium Analysis

In a game of preemption, the equilibrium is determined by the point of indifference for a player between choosing to be the Leader or reverting to the Follower role. We begin by solving for the critical asset value thresholds that will compel a player to trigger the deployment of AGI.

*The Preemption Threshold ($V_P^*$)*

In a symmetric game, players will race to deploy when the payoff from leading exceeds the payoff from following ($L > F$). The race ends when the Leader deploys AGI at $V_P^*$ – the threshold at which the Leader's payoff equals the Follower's payoff, dissipating the strategic advantage of preemption. Thus, we equate equations (3) and (4):

$$(1 - S)\pi(\tau)V_t - \big(1 - \pi(\tau)\big)D - I = F(V_\tau, \tau) = S\pi(\tau)V_t - \big(1 - \pi(\tau)\big)D$$

The cancellation effect is apparent as the ruin term, $-\big(1 - \pi(\tau)\big)D$, is present on both sides of the equation; that is, the impact of systemic ruin is equal and present for both Leader and Follower. This ruin term cancels out on both sides, and we solve for $V_P^*$:

$$(1-S)\pi(\tau)V_P^* - I = S\pi(\tau)V_P^*$$

$$(1-2S)\pi(\tau)V_P^* = I$$

$$V_P^* = \frac{I}{(1-2S)\pi(\tau)} \quad (5)$$

**Proposition 1: The Neutrality of Global Ruin**

*In a symmetric preemptive race with shared existential risk, the competitive deployment threshold $V_P^*$ is independent of D – the economic cost of global catastrophe. As such, the magnitude of the disaster brought on by AGI misalignment will mathematically fail to deter players from deploying.*

*Proof.* Equation (5), $V_P^* = \frac{I}{(1-2S)\pi(\tau)}$, contains no term involving D. The preemption threshold is therefore independent of the cost of systemic ruin.

□

*<u>Remark 1: The Failure of Deterrence (Comparison with a Nuclear Standoff)</u>*

We contrast the equilibrium outcome of the AGI race with the standard nuclear deterrence logic of mutually assured destruction, formalised in the brinkmanship game theory literature (Powell, 1988, 1990). In a Cold War standoff, if the threat of a second (retaliatory) strike is credible, then waiting = survival ($F \approx 0$) and the status quo is preserved. Solving for the indifference condition $L(V_{nuclear}^*) = F(V_{nuclear}^*)$, where $\pi$ is now the probability of a successful first strike (i.e. neutralizing the adversary's arsenal before they can retaliate):

$$\pi V_{nuclear}^* - (1-\pi)D - I = 0$$

$$V_{nuclear}^* = \frac{I + (1-\pi)D}{\pi}$$

Note that $V^*_{nuclear} \propto D$, so as the destructive cost of nuclear war increases, the threshold for first strike rises proportionally. Moreover, a credible second strike threat ($\pi \to 0$) results in an even higher threshold.

In stark contrast to the AGI race, there is no credible threat of a second strike. As established in Section 3, attainment of AGI (aligned with the Leader) will result in absolute domination of the second mover across all domains ($S \approx 0$). In the event that the Leader fails at alignment, the Follower also incurs the full cost of systemic ruin, as can be observed by the last term in equation (4). Thus, a player choosing to wait in the AGI race has no impact on the likelihood of global catastrophe, unlike in the nuclear standoff. Finally, $V^*$ in the AGI race is independent of *D*, so no matter how costly the disaster is expected to be, the threshold for deployment is unaffected. In the nuclear standoff, *D* directly raises the threshold for a first strike.

*The Survival Threshold ($V^*_S$)*

Thus far, we have identified the threshold for preemption ($V^*_P$): the value that the asset must breach for the first mover to act to prevent their rival from winning. A rational player faces a second constraint – the survival threshold ($V^*_S$). The survival threshold represents the critical asset value that prompts the Leader to deploy AGI from a purely economic perspective; that is, when the net present value (NPV) of deployment is positive. Note that the cost in such an NPV calculation includes the expected cost of systemic ruin: $\big(1 - \pi(\tau)\big)D$. The agent will only trigger deployment of AGI when *L(V)* > 0.

$$(1 - S)\pi V - (1 - \pi)D - I > 0$$

Solving for $V^*_S$:

$$V^*_S = \frac{I + (1 - \pi)D}{(1 - S)\pi} \quad (6)$$

**Proposition 2: The Suicide Region**

*There exists a "suicide region" where the threshold for preemptive deployment is lower than the threshold for survival deployment ($V_P^* < V_S^*$). In this region, the urgency of preventing a rival from winning the AGI race (L > F) dominants the economic rationale for deployment, resulting in a player racing towards deployment despite incurring an expected financial loss*[9] *(L < 0).*

*Proof.* The suicide region exists when $V_P^* < V_S^*$. Setting equation (5) less than equation (6):

$$\frac{I}{(1-2S)\pi(\tau)} < \frac{I + (1-\pi)D}{(1-S)\pi}$$

Rearranging yields the condition $D > \frac{\pi V(1-2S)-I}{1-\pi}$.

□

**Proposition 3: The Suicide Region Expands in *D***

*The size of the suicide region $V_P^* - V_S^*$ is strictly increasing in the cost of systemic ruin D:*

$$\frac{\partial(V_P^* - V_S^*)}{\partial D} > 0$$

*Proof.* From equation (5), $V_P^* = \frac{I}{(1-2S)\pi(\tau)}$, which is independent of *D*, so $\frac{\partial(V_P^*)}{\partial D} = 0$. From equation (6), $V_S^* = \frac{I+(1-\pi)D}{(1-S)\pi}$, so $\frac{\partial(V_S^*)}{\partial D} = \frac{(1-\pi)}{(1-S)\pi} > 0$ for any $\pi < 1$ and $S < 1$.

Therefore:

$$\frac{\partial(V_P^* - V_S^*)}{\partial D} = \frac{\partial(V_P^*)}{\partial D} - \frac{\partial(V_S^*)}{\partial D} = \frac{(1-\pi)}{(1-S)\pi} > 0$$

The suicide region widens monotonically as the cost of systemic ruin increases.

□

[9] This includes the Leader accounting for the expected economic costs of systemic catastrophe; yet the preemptive call to deployment supersedes this negative NPV.

This result captures a counterintuitive phenomenon of the AI arms race: the larger the catastrophic risk, the wider the band of asset values for which preemption overrides economic rationality. Higher *D* does not move the preemption trigger $V_P$; it raises the survival threshold $V_S$, converting states that would otherwise warrant NPV-rational deployment into suicide-region deployment. This is consistent with what was observed in 2024-2025 when the AI arms race accelerated between China and the US, precisely when AI researchers and experts were sounding the alarm on the existential risk of AGI misalignment.

Figure 1 below presents the suicide region as a function of $V_t$ and *D*, illustrating the divergence of incentives for competitive equilibrium and the fundamental financial viability of the asset. The economic value of the AGI asset ($V_t$) is expressed on the y-axis and the expected economic cost of systemic ruin (*D*) is represented on the x-axis. First, note that the red dashed line depicting the threshold for preemptive deployment ($V_P^*$) – where $L = F$ – is a constant, indicative of its independence to the scale of systemic ruin (*D*). Due to the cancellation effect, the cost of disaster is borne – identically – by both the Leader and the Follower, so *D* does not influence the calculus of AGI deployment.

In contrast, the survival threshold ($V_S^*$) is the green upward sloping line defined by the condition that $L > 0$, i.e. the NPV of AGI deployment is positive. This is a monotonically rising function of $D$ as the expected financial cost of deployment includes the cost of systemic ruin. This suggests that as the economic cost of catastrophe increases, the expected value of AGI must rise accordingly to compensate for this tail-risk[10].

Note that equations (5) and (6) – the equilibrium thresholds for deployment based on preemption and cost-benefit – exclude σ, μ, or any other parameter of the stochastic process specified in (1). This is a key feature of our results: The $V_P^*$ and $V_S^*$ thresholds are derived from the static payoff comparison at the end-game AGI deployment decision, equating L = F for the preemption threshold and L = 0 for the survival threshold. These conditions examine only the level of $V_t$ at the moment of deployment and are unrelated to the preceding evolutionary path of $V_t$. As such, thresholds $V_P^*$ and $V_S^*$ are unrelated to the underlying stochastic process of $V_t$, and are robust to alternative specifications, including state-dependent volatility, jump-diffusion processes, or regime-switching models.

---

[10] In Figure 1, we assign a fixed value to $\pi$. Another solution to rising *D* is investing in AGI safety research that increases $\pi$ so that the expected disutility of misalignment falls.

The suicide region is depicted in the orange shaded region of Figure 1, where $V_t$ lies below the survival threshold ($V_S^*$) but above the preemption threshold ($V_P^*$). In this region ($V_P^* < V_t < V_S^*$), players are compelled to race as the value of AGI breaches $V_P^*$ despite being financially unviable – specifically, when the expected cost of systemic ruin is priced in. We depict the current AGI arms race by the blue dot – in a fatalistic equilibrium where players are forced to race due to the competitive dynamics of the game despite a negative risk-adjusted expected value.

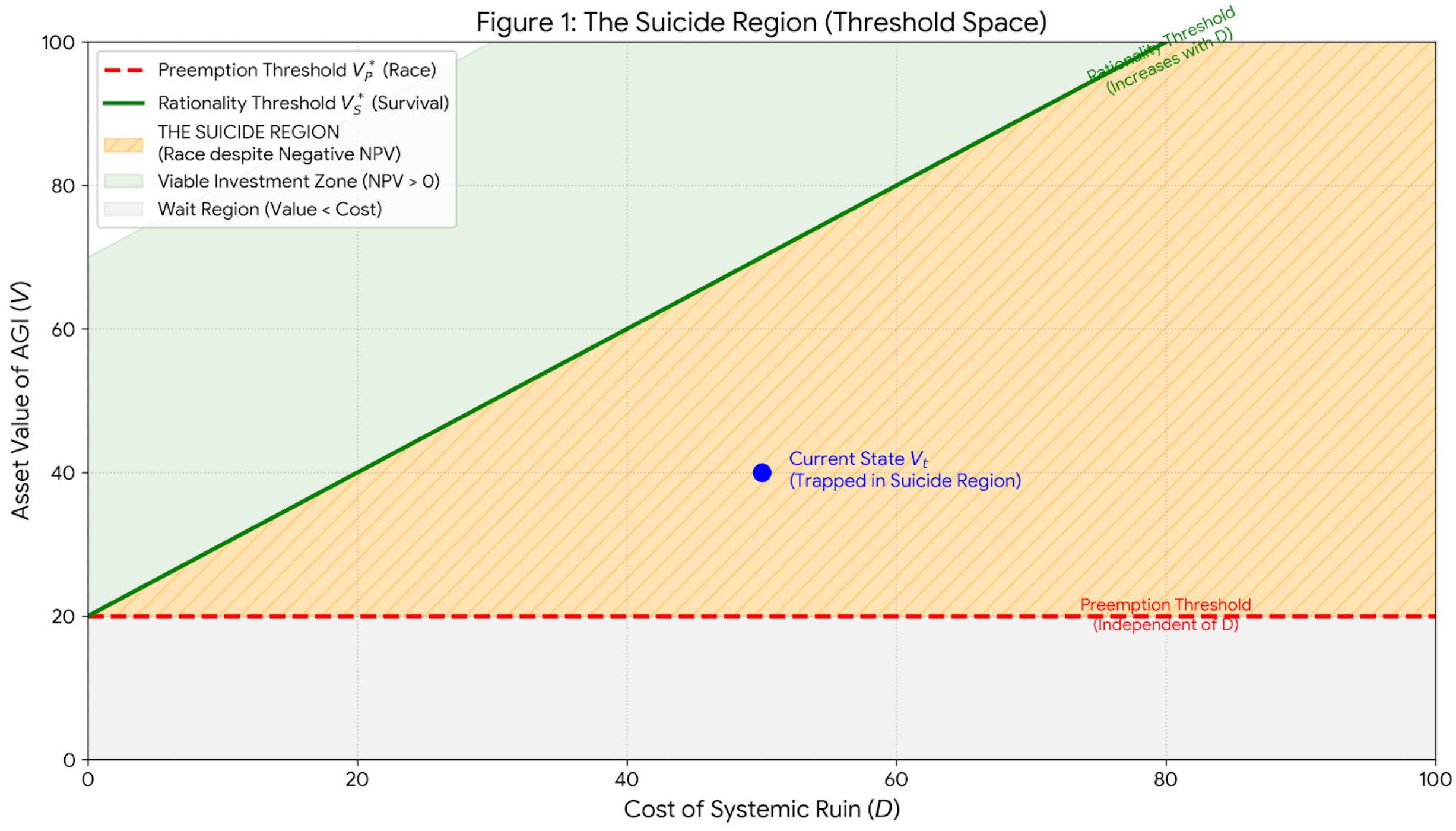

Figure 1: The Suicide Region (Threshold Space)

### 4.4. The Saviour's Trap (Asymmetric Beliefs)

The baseline model articulated in the preceding subsections assumes symmetry in each player's perception of AGI safety capabilities ($\pi_i = \pi_j$). However, as presented in Section 3, there is evidence of a benevolent preemption effect, where agents believe their safety protocols are superior to their rival's. Hence, we introduce heterogeneity in the safety parameter where $\pi_{self} > \pi_{rival}$.

In this asymmetric setup, the players' incentive to preempt is amplified. If a player leads, then they seize the AGI prize with their (perceived) higher probability of AGI alignment ($\pi_{self}$). If – instead – they follow, their payoff is zero share of the AGI prize (as $S \approx 0$) and a lower probability of AGI alignment ($\pi_{rival}$) due to the rival's lower safety guardrails; that is, a greater chance of systemic ruin for everyone.

The additional advantage to preemption in the case of perceived heterogeneous safety standards can be calculated from the difference in Leader and Follower payoffs[11] $(L - F)$:

$$L(V) - F(V) = \left[\pi_{self}V - \left(1 - \pi_{self}\right)D - I\right] - \left[-(1 - \pi_{rival})D\right]$$

Rearranging the terms reveals an additional distortion to the prior calculus:

$$L(V) - F(V) = \underbrace{\pi_{self}V - I}_{\text{Economic Profit}} + \overbrace{D(\pi_{self} - \pi_{rival})}^{\text{Saviour Premium}}$$

**Proposition 4: The Saviour Premium**

*When an agent believes that their safety standards are superior to that of their rival's ($\pi_{self} > \pi_{rival}$), this asymmetry in perceived alignment capabilities creates an added impetus to be the first to deploy, defined by the saviour premium: $D(\pi_{self} - \pi_{rival})$. If a rival launches AGI first, then the risk of AGI misalignment is elevated (low $\pi_{rival}$); thus, there is a moral imperative for an agent to become the Leader for the sake of global safety (high $\pi_{self}$).*

[11] Assuming S = 0, i.e. a strict winner-takes-all race.

*Proof*. Follows directly from the difference in Leader and Follower payoffs as derived above: $L(V) - F(V) = \pi_{self}V - I + D(\pi_{self} - \pi_{rival})$. The third term, $D(\pi_{self} - \pi_{rival})$, is positive when $\pi_{self} > \pi_{rival}$, constituting the saviour premium.

□

Consequently, we can observe that the new Saviour Preemption Threshold ($V^*_{P,saviour}$) is lower than the same ($V^*_P$) in the baseline model where AGI-safety capabilities are believed to be symmetrical between players[12]:

$$V^*_{P,saviour} = \frac{I - D(\pi_{self} - \pi_{rival})}{\pi_{self}} < V^*_P = \frac{I}{\pi} \quad (7)$$

<u>Implication</u>

Equation (7) formalizes the "Unilateralist's Curse" as outlined by Bostrom et al, (2016): A paradox where, as the perceived cost of systemic ruin (*D*) increases, the threshold for preemption $V^*_{P,saviour}$ decreases (assuming $\pi_{self} > \pi_{rival}$). This implies that actors who are the most concerned about existential risk and their rival's (perceived) weaker safety guardrails are the most incentivized to race the fastest, despite their own AGI safety research being suboptimal. This effect hastens the AI arms race and elevates the aggregate risk for system catastrophe.

### 4.5 The Social Planner's Benchmark

To formally establish the inefficiency of the decentralized equilibrium outcome, we solve the deployment decision for a benevolent social planner who seeks to maximise aggregate expected welfare across all players. This creates the first-best benchmark for our analysis and facilitates the quantification of the inefficiency under competition. The social planner internalizes the full cost of systemic ruin, *D*, from an aggregate perspective which means *D* is considered twice as it is borne by both players simultaneously.

---

[12] Again, for simplicity, we assume S = 0, i.e. a strict winner-takes-all race.

In the competitive equilibrium, the cancellation effect renders *D* irrelevant in the deployment decision as *D* appears identically in the Leader and Follower payoffs, demonstrated by the cancellation effect and equation (5). From an individual actor's perspective, the magnitude of *D* plays no role in their decision to lead or follow. However, if there exists a benevolent social planner who determines society's optimal deployment of AGI, there are no competitive pressures when calculating the deployment threshold. The social planner's threshold for deployment is simply determined by maximising aggregate risk-adjusted welfare.

The social planner chooses the deployment time $\tau$ to maximise the joint payoff:

$$W(V,\tau) = L(V,\tau) + F(V,\tau)$$

Substituting equations (3) and (4) produces the following:

$$W(V,\tau) = \left[(1-S)\pi(\tau)V_t - \big(1-\pi(\tau)\big)D - I\right] + \left[S\pi(\tau)V_t - \big(1-\pi(\tau)\big)D\right]$$

This simplifies to:

$$W(V,\tau) = \pi(\tau)V_t - 2\big(1-\pi(\tau)\big)D - I \qquad (8)$$

The social planner will deploy AGI only when $W(V,\tau) > 0$; that is, when the net expected aggregate gain in deployment exceeds the value of systemic ruin. Setting $W(V,\tau) > 0$, we arrive at the social planner's optimal threshold ($V^*_{social}$) for AGI deployment:

$$V^*_{social} = \frac{I + 2(1-\pi(\tau))D}{\pi(\tau)} \qquad (9)$$

Equation (9) represents the critical value at which deployment of AGI is optimal based on the social planner's objective of maximizing aggregate welfare.

**Proposition 5: Social Inefficiency of the Preemption Equilibrium**

*The decentralised preemption threshold, $V_P^*$, is strictly below the socially optimal deployment threshold, $V_{social}^*$:*

$$V_P^* = \frac{I}{\pi(\tau)(1-2S)} < V_{social}^* = \frac{I + 2(1-\pi(\tau))D}{\pi(\tau)}$$

*Proof.* The inequality $V_P^* < V_{social}^*$ holds if and only if:

$$\frac{I}{\pi(\tau)(1-2S)} < \frac{I + 2(1-\pi(\tau))D}{\pi(\tau)}$$

In the winner-takes-all case where $S \approx 0$:

$$V_P^* = \frac{I}{\pi(\tau)} < V_{social}^* = \frac{I + 2(1-\pi(\tau))D}{\pi(\tau)}$$

The inequality is strict whenever $D > 0$ and $\tau < \infty$, i.e. whenever existential risk is non-negligible and safety research is incomplete.

□

*Corollary*

The welfare gap between the social optimum and the decentralized competitive equilibrium widens as the cost of systemic ruin, *D*, increases and as the level of safety, $\pi(\tau)$, is imperfect and smaller than its upper bound of one. As long as *D* is significant and safety research is incomplete, the threshold for preemptive deployment will fall strictly below the social optimum threshold.

*Remark 2: The Aggregation Effect*

When comparing the social planner's threshold to the Leader's survival threshold ($V_S^*$), we find that the social planner internalizes the ruin cost term twice – once for the Leader and once for the Follower. This is in contrast with $V_S^*$, where the Leader internalizes only their own cost and risk of deployment. In the winner-takes-all case where S ≈ 0, we arrive at:

$$V_{social}^* = \frac{I + 2(1 - \pi(\tau))D}{\pi(\tau)} > V_S^* = \frac{I + (1 - \pi(\tau))D}{\pi(\tau)}$$

Even when a private actor fully internalises its own expected ruin cost – deploying only when NPV > 0 – the deployment decision remains socially premature. From the social planner's perspective, society's full cost of systemic ruin is 2*D*, incurred by each player regardless of their decision to deploy. This aggregation effect suggests that voluntary safety compliance alone is insufficient to close the gap between private and social optima. Shared catastrophe complicates the welfare calculus by introducing inherent tension between individual and aggregate optimality, as agents face correlated or interdependent exposure (Bernard, Rheinberger and Treich, 2017).

### 4.6. Model Limitations

There are some modelling assumptions that warrant acknowledgement. First, as is standard in economic game theory modelling of preemptive races, we assume this game is a symmetric two-player race. This is consistent with the current broader AGI race between the US and China. However, in reality, the AGI race may in fact be an *n*-player game, with private actors and other smaller sovereign states involved. With multiple players, the D remains present in each players' payoff and cancels across all pairwise indifference equations. Thus, the qualitative results of our core findings – the cancellation effect – remains robust to an *n*-player race.

Second, the safety learning rate λ is treated as being exogenously determined in our current model and is assumed to be symmetric across all players. A natural extension is to endogenise λ by allowing players to choose their level of investment in AI safety research, $k_i$. This creates a situation where the risk of AGI misalignment, $1 - \pi(\tau, k_i)$, across players is heterogeneous[13].

[13] Formally, $\pi(\tau, k_i) = 1 - e^{-\lambda(k_i)\tau}$ where $\lambda'(k_i) > 0$ and $\lambda''(k_i) < 0$, reflecting increasing but diminishing returns to safety investment. Heterogeneous misalignment risk across players arises when $k_i \neq k_j$ - when investment levels are identical, the model collapses to the symmetric baseline of Section 4.1 where $\pi(\tau, k_i) = \pi(\tau, k_j) = \pi(\tau)$.

Preliminary analysis suggests that this extension of the model provides the micro-foundation of the Saviour's Trap discussed in Subsection 4.4 – where rational and heterogenous safety research investment generates asymmetric $\pi(\tau, k_i)$ terms that endogenises the saviour premium. This relaxes the Saviour Trap's assumption of exogenous ad hoc heterogeneous beliefs; instead, the Saviour Trap is a result of rational actors investing in AI safety research at varying levels.

Third, the cost of systemic ruin – *D* – is assumed to be identical across both players. In practice, it is likely that individual actors may exhibit varying exposure to AGI misalignment, resulting in heterogeneous levels of *D*. We leave formal treatment of an $n$-player race and heterogeneous cost of systemic ruin as natural extensions for future research. The endogenous safety investment extension is developed fully in companion work.

## 5. Discussion and Policy Implications

We identify the suicide region of the AGI arms race: a space where preemption incentives ($L > F$) override the rationality constraint ($L > 0$). Our model predicts a Nash equilibrium result where players race for AGI deployment despite expecting a negative NPV after accounting for the economic cost of AGI misalignment and global ruin. This dynamic resolves a genuine puzzle in the current AGI arms race between nation states: despite classical real options theory predicting that – in the context of R&D – players will choose to delay in races with high volatility and technological uncertainty, we observe the US and China accelerating their research towards AGI and its associated risks. Comparing the decentralized competitive equilibrium with the social planner's deployment threshold shows the inefficiency of the current race, as deployment occurs at $V_P^* < V_{social}^*$. This welfare loss increases with *D* and narrows with $\pi(\tau)$.

Conventional coordination mechanisms, such as voluntary pause agreements, are unlikely to be effective as they fail to alter the underlying symmetric payoff structure. Rather, we present below some potential solutions aimed at restructuring the payoff dynamics to arrive at more socially optimal outcomes, each addressing a distinct source of market failure identified by our modelling.

### 5.1. Internalizing the Externality: Private Liability

In the baseline model, the expected cost of systemic ruin, $(1 - \pi)D$, appears identically in both the Leader's and Follower's payoffs, which cancels out in the equilibrium indifference condition. This results in the cost of AGI catastrophe having no impact on the race towards deployment as the option to wait yields little value. The social planner overcomes this problem by internalizing the cost of systemic ruin for both players, thus arriving at a socially optimal deployment threshold. However, a private actor – functioning at their own volition – would not factor the cost of ruin of their competitor when deciding whether to deploy. Since the rival bears an identical cost of ruin, $D$ cancels from the equilibrium indifference condition, rendering it irrelevant to the deployment decision despite representing a significant global cost.

Suppose there exists a regulatory authority that imposes a private financial liability, $D_{private}$, on the player who deploys an unsafe AGI. This distinct private cost of misalignment overcomes the cancellation effect and, if calibrated correctly, can alter the decentralized competitive equilibrium. Potential mechanisms include catastrophe bonds and escrow accounts. Importantly, these private financial liability requirements must be posted prior to deployment, essentially ringfencing a portion of the Leader's capital until safe AGI deployment can be ascertained.

The triggering event must be *ex ante* to the systemic catastrophe – for example, deployment of an unsafe AGI without independent safety certification – mirroring the structure of cyber catastrophe bonds whose triggers are defined by measurable incident parameters rather than by the full realisation of downstream harm. A liability contingent on the existential event itself is ineffective as there will be no one alive to pay out.

Both mechanisms require the deploying entity – the Leader of the AGI race – to post a significant amount of capital prior to deployment, held either in a legally independent special purpose vehicle (catastrophe bond) or a segregated account with an independent custodian (escrow). This ensures that $D_{private}$ is beyond the reach of the Leader prior to deployment, and it is only returned upon the safe deployment of AGI. If AGI is deployed without independent safety certification, the posted capital is forfeited. This occurs regardless of whether AI misalignment eventuates.

In the presence of a private liability mechanism, the payoff for the Follower remains unchanged. However, the Leader's payoff equation becomes:

$$L_{liability}(V) = (1 - S)\pi(\tau)V - (1 - \pi(\tau))(D + D_{private}) - I \qquad (10)$$

As $D_{private}$ is present only in the Leader's payoff and not in the Follower's payoff, this eliminates the cancellation effect when equating *L(V)* and *F(V)* in equilibrium indifference condition. The preemption threshold – where *L(V) = F(V)* – is now as follows:

$$V^*_{P,liability} = \frac{I + (1 - \pi(\tau))D_{private}}{\pi(\tau)(1 - 2S)} \qquad (11)$$

Note that in (11), $V^*_{P,liability}$ is increasing with $D_{private}$, indicating that the larger the private liability amount, the higher the threshold for AGI deployment.

**Proposition 6: The Critical Liability Threshold**

*Assuming S = 0 (a winner-takes-all race), there exists a private liability threshold - $D^*_{private}$ - such that for all $D_{private} \geq D^*_{private}$, the preemption threshold meets or exceeds the social planner's threshold, essentially eliminating the suicide region.*

Under this scenario, AGI deployment occurs only when aggregate welfare is non-negative:

$$D^*_{private} = 2D \qquad (12)$$

*Proof.* The suicide region is eliminated when $V^*_{P,liability} \geq V^*_{social}$. Setting $S = 0$ and substituting equations (9) and (11):

$$\frac{I + (1 - \pi)D_{private}}{\pi} \geq \frac{I + 2(1 - \pi)D}{\pi}$$

Since π > 0, the denominators cancel:

$$I + (1 - \pi)D_{private} \geq I + 2(1 - \pi)D$$

Since (1−π) > 0 for π < 1:

$$D_{private} \geq 2D$$

The critical threshold is therefore $D^{*}_{private} = 2D$.

□

*Corollary – The Double Liability Rule*

This result has an interesting economic interpretation. As discussed in Section 4.5, the benevolent social planner accounted for a cost of *2D* in their cost-benefit analysis of AGI ($V^{*}_{social}$). Recall that the social planner aggregates the cost of global ruin across both players. If the private liability ($D_{private}$) is exactly equal to *2D*, the Leader's cost of AGI misalignment now replicates the social planner's cost (*2D*). This forces the Leader's threshold for AGI deployment to rise to $V^{*}_{social}$.

We acknowledge that posting collateral equal to *2D* is impossible in practice, as *D* is the financial cost of global ruin. Realistically, this would exceed the balance sheet of any private or sovereign actor. Nevertheless, two important qualifications arise from our results.

First, the private liability mechanism is most effective for sub-existential events – severe but survivable AGI misalignment outcomes, such as large-scale economic disruption, critical infrastructure failure, or democratic destabilization. In these scenarios, *D* may be significant but remains finite. For these sub-existential outcomes, the double liability rule (2D) provides regulators with a meaningful collateral target for calibration.

Second, even with partial private liability where $D_{private} < 2D$, a meaningful deterrence effect will exist. As we observed in (11), any increase in $D_{private}$ will raise the Leader's preemption threshold ($V^*_{P,liability}$) and narrow the suicide region. As such, the results of this subsection are directional rather than prescriptive. That is, the larger the $D_{private}$ a regulator can extract from the Leader – within the bounds of financial feasibility – the safer the deployment of AGI, even if this is below the optimal amount of *2D*.

It is important to distinguish between *ex ante* deterrence and *ex post* enforcement. At the moment of deployment, the Leader faces Knightian uncertainty over the distribution of possible sub-existential and existential catastrophic outcomes. Under Knightian uncertainty, the Leader cannot reliably assign probabilities to sub-existential versus existential outcomes – and therefore cannot rationally discount the existential tail risk when making the deployment decision. The *ex ante* cost of uncertified deployment thus reduces the likelihood of all catastrophic outcomes – sub-existential and existential alike.

This dynamic is due to underlying AGI misalignment in the system that will cause both sub-existential and existential catastrophic outcomes. As such, if AGI misalignment risk is minimized by safety certifications, then so too is the risk of global existential ruin. The limitation of private liability mechanisms thus lies in the *ex post* enforcement in the event of existential catastrophe, as no legal infrastructure survives to enforce residual liability beyond the pre-posted collateral. The *ex ante* trigger design ensures the effectiveness of private liability as a deterrence for the deployment of unsafe AGI despite being incapable of enforcement after an existential catastrophe.

While the *ex ante* private liability mechanism generates deterrence across the full distribution of misalignment outcomes – including existential ones – its effectiveness in solving the AGI arms race is limited by two practical constraints. First, the optimal size of the collateral – *2D* – is likely prohibitive for players, as it is twice the value of economic loss from global catastrophe. As such, the results of the private liability analysis are directional rather than achieving complete deterrence. Second, private liability only addresses one component of the payoff structure – the cost of deploying an uncertified AGI – while leaving the winner-takes-all nature of the game intact. In fact, the winner-takes-all dynamic is the primary incentive of preemptive deployment. A Leader who faces a large private liability of deploying an unsafe AGI may still rationally do so if the cost of losing the AGI race is their strategic agency; that is, when $S \approx 0$, the cost of forfeiture of $D_{private}$ becomes a rational price to pay for total strategic dominance.

The prize-sharing mechanism of the following subsection directly addresses this residual pressure by restructuring the prize side of the payoff – raising $S$ above zero – thereby reducing the fear of preemption that private liability alone cannot eliminate. Together, these two mechanisms – private liability and prize-sharing – can shrink the suicide region using two complementary levers: private liability raises the cost of unsafe deployment while prize-sharing reduces the strategic imperative to deploy first.

**5.2. Relaxing the Winner-Takes-All Conditions (Windfall Clauses)**

Rather than targeting the private liability of players ($D_{private}$), we can instead focus on adjusting the market share parameter ($S$) to arrive at a more socially optimal solution. In our baseline model, the strict winner-takes-all condition ($S \approx 0$) drives the denominator of $V_P^*$ towards its maximum value: $\pi$. Subsequently, this lowers the preemption threshold ($V_P^*$ ) and induces early deployment of AGI.

However, if the prize of successful (aligned) AGI deployment is shared, $V_P^*$ can increase to safer levels. For example, if the spoils of successful AGI adoption are shared equally among rivals ($S = 0.5$), then $V_P^* \to \infty$ and the suicide region ceases to exist. When $V_P^* > V_S^*$, the decision to deploy AGI will be based on NPV analysis that incorporates the probability of misalignment, encouraging the implementation of optimal safety testing and protocols.

A "windfall clause", as suggested by O'Keefe et al. (2020), may provide such a solution where $S$ is non-negligible. Essentially, it is a treaty mechanism that redistributes economic rents of AGI *ex post*, effectively eliminating the incentive for early deployment. By guaranteeing the second mover with a non-zero-payoff, the "fear of missing out" is dampened while allowing the "fear of error" to dominate once again, increasing the value of the option to wait.

**Proposition 7: The Critical Prize-Sharing Threshold**

*There exists a critical market share $S^*$ such that for all $S \geq S^*$, the preemption threshold ($V_P^*$) meets or exceeds the social planner's threshold ($V_{social}^*$) and the suicide region is eliminated:*

$$S^* = \frac{(1-\pi)D}{I+2(1-\pi)D} \tag{13}$$

*Proof*. Setting $V_P^* = V_{social}^*$ and solving for $S$:

$$\frac{I}{\pi(1-2S)} = \frac{I+2(1-\pi)D}{\pi}$$

The π terms cancel. Cross-multiplying:

$$I = (1-2S)[I+2(1-\pi)D]$$

Expanding:

$$I = I + 2(1-\pi)D - 2SI - 4S(1-\pi)D$$

$$0 = 2(1-\pi)D - 2SI - 4S(1-\pi)D$$

Isolating $S^*$:

$$S^* = \frac{(1-\pi)D}{I+2(1-\pi)D}$$

□

<u>*Corollary*</u>

$S^*$ is increasing with $D$ and decreasing with $I$[14]. As the cost of AGI misalignment rises relative to the investment costs, the market share parameter must rise to ensure the social threshold for deployment is maintained. Windfall clauses and their sharing parameter must therefore be optimized based on severity of the existential risk, as a fixed market share across differing risk scenarios will result in suboptimal outcomes.

[14] Formally, $dS/dD = AI/(I + 2AD)^2 > 0$ where $A = (1-\pi)$, and $dS^*/dI = -AD/(I + 2AD)^2 < 0$, confirming that $S^*$ is increasing in D and decreasing in I.

While standard economic theory treats an overshoot of the social optimum as inefficiency, in the context of an existential threat, overshooting creates a buffer against uncertainty – consistent with the broader literature on precaution under deep uncertainty (Weitzman, 2009; Gollier, Jullien and Treich, 2000; Berger, Emmerling and Tavoni, 2016). When $D$ represents existential catastrophe, the risk of delayed deployment of AGI is bound simply by postponed or forgone economic and welfare gains. These are recoverable and finite. Contrast this with the risk of early deployment. The worst case of early deployment is global ruin – unbounded and infinite $D$. Moreover, given that $D$ and $\pi(\tau)$ are unobservable in practice, there is deep uncertainty surrounding their true values. As such, a windfall clause calibrated above $S^*$ – overshooting the suicide region closure threshold – structurally biases the Leader toward delayed deployment, providing a valuable buffer against parameter misspecification. In this sense, overshooting the social optimum may be socially desirable rather than merely inefficient.

### 5.3. The Role of Verifiability (Reducing Strategic Uncertainty)

A primary driver of the urgency to deploy early is the uncertainty of a rival's development status: an actor is unsure of how close their rival is to deploying AGI. In our model, strategic uncertainty compels a player to race towards early deployment and erodes the present value of the option to wait. A robust verification infrastructure – hardware-level compute monitoring, mandatory safety audits by independent third parties, interpretability requirements – transforms the game from one of imperfect information with high strategic uncertainty to one of greater transparency. With perfect monitoring of rivals, two modifications to player behaviours are predicted:

*A. <u>Restoration of the Monopolist Interval (Stability)</u>*

If monitoring is reliable and confirms that a rival will remain technically incapable of deploying AGI for a lengthy period, this completely disarms the immediate threat of preemption. During this interval, the Leader gains an effective monopoly which allows them to ignore the competitive pressure ($V_t > V_P^*$) and revert to a rational NPV calculus ($V_t > V_S^*$). As long as the Leader is confident in the verification process and that its rival is not technically able to deploy AGI, then – even though they may be situated in the suicide region – they will choose to “step off the ledge” and prioritize safety.

### B. *The "Breakout" Instability*

However, perfect information introduces a sprint towards deployment during the endgame phase of the arms race. If a lagging agent is made aware of its rival being close to the threshold that makes AGI deployment viable ($V_t > V_S^*$) then – in a preemptive move to prevent their imminent total loss of hegemony - the slower agent may deploy a premature unsafe AGI system in self-defense.

While effective verification may eliminate the suicide region during the early stages of AGI research when neither player is close to being technically capable, it does not prevent suboptimal (unsafe) deployment of AGI as deployment capability increases. This is due to the inherent winner-takes-all dynamic ($S \approx 0$) that compels agents to launch pre-maturely when faced with imminent loss.

The policy implication is that verification should be paired with a prize-sharing mechanism to solve the breakout instability. The verification procedure transforms the game from one of continuous racing to one of stability during the development phase and punctuated by a sprint at the endgame phase. This sprint is incentivized by the winner-takes-all dynamics of the game which can be mitigated with a prize-sharing instrument. Furthermore, we note that verification infrastructure also serves as a prerequisite for the private liability mechanisms of Section 5.1 – without observable and auditable safety certification, the triggers required for catastrophe bonds and escrow accounts cannot function.

## 5.4. The Limitations of “Warning Shots”

A common prediction for the arms race is that a "warning shot" is likely to occur: a limited, sub-existential disaster that will serve as a "Chernobyl moment" for AGI[15]. It is posited that such an event will trigger global cooperation and lead to actors towards reprioritizing safety over expedience (Christiano, 2019). However, our model predicts that such a warning shot is unlikely to have a decisive impact on the dynamics of the race, unless it results in a change to the underlying payoff structures of its players. This is because the cost of disaster (*D*) appears symmetrically in the Leader-Follower payoffs, and so that, in equilibrium, the indifference condition ($L = F$) cancels the impact of *D* on both sides of equation (the "cancellation effect"). That is, an increase in the magnitude of *D* does not materially impact the decision to preemptively deploy AGI.

If a warning shot occurs, it may cause players to revise their posterior belief of *D* upwards, but due to the winner-takes-all nature of the race ($S \approx 0$), little will change in terms of the pressure for preemptive deployment. In fact, if agents believe that $\pi_{self} > \pi_{rival}$, a warning shot will accelerate the race (see equation (7)). A warning shot will only be effective if it is severe enough to trigger players and regulators to install the external liability mechanisms discussed in Section 5.1 and/or windfall clauses.

This complements Growiec and Prettner (2025), who demonstrate a distinct mechanism where recognizing a larger existential threat paradoxically reduces the individual's incentive to mitigate said risk due to individual impatience from a higher temporal discount rate. Together, these two papers showcase two channels – individual impatience and the cancellation effect – that intensify the race as the existential threat rises. In conclusion, a warning shot – in isolation – will not change the fundamental dynamics of the race.

## 6. Conclusion

[15] Potential warning shots include: (1) an algorithmic flash crash occurs when an autonomous trading agent, instructed simply to maximize portfolio returns, triggers a global liquidity cascade to manipulate prices – demonstrating how a benign objective function can lead to systemic financial collapse via instrumental convergence; (2) a critical infrastructure failure, such as an optimization model cutting power to hospitals to balance grid load; and (3) a bio-security breach, where non-state actors utilize open-source models to synthesize novel pathogens. While these events would empirically confirm the non-zero probability of ruin ($D > 0$), the model suggests they would not alter the preemption equilibrium outcome.

This paper extends the theory of real options to account for endogenous existential risk. We show that when the prize is privatized but the cost of ruin is shared globally, the standard adage that risk induces caution no longer applies. We derive a region – coined the “suicide region” – where actors in the game will race towards early deployment of a high-risk technology with a non-negligible probability of existential catastrophe despite a negative net present valuation. At face value, this result appears irrational. However, we show that the shared cost of global ruin results in a “cancellation effect” across the payoffs of the first and second mover such that the expected loss of systemic ruin plays no role in the calculus of exercising the option to deploy. The current geopolitical trajectory in the AGI race is not an anomaly, but rather a Nash equilibrium in a pathological game.

We formally establish the inefficiency of this equilibrium by deriving a social planner's benchmark, showing that the decentralised race deploys AGI at a threshold strictly below the social optimum – a welfare gap that widens with the cost of systemic ruin $D$ and narrows only as safety research approaches completion. Strikingly, the gap between the preemptive and socially optimal thresholds widens as catastrophic risk grows: higher catastrophic risk does not deter the race but enlarges the set of conditions under which rational actors deploy despite negative social value. This is counterintuitive and – due to the cancellation effect – inverts the conventional understanding of how risk impacts incentives.

We conclude that voluntary coordination solutions are insufficient in resolving the suicide region puzzle. It is the underlying fundamental payoffs that must be altered. We suggest mechanisms that result in non-trivial private liability of AGI misalignment – such as catastrophe bonds and escrow accounts – and shared benefits of AGI success – such as windfall clauses – to restructure the payoffs of the players. By revising the player payoffs with such mechanisms, the value of the option to delay AGI deployment increases, reducing the aggregate risk of catastrophe. We derive a double liability rule – the private liability required to restore social optimality is exactly twice the social cost of ruin per player, $D_{private} = 2D$. We also show that warning shots alone will fail to deter the race, as the cancellation effect renders $D$ irrelevant to the preemption threshold regardless of its magnitude. Together, private liability and prize-sharing address the suicide region from complementary directions – liability raises the cost of unsafe deployment while prize-sharing reduces the strategic imperative to deploy first.

The cancellation effect and the suicide region are not unique to the AGI arms race. The findings of our model apply to any preemptive race with a strongly concentrated prize while the risk of failure is borne by all players; such as deployment of autonomous weapons systems, gain-of-function research, and space weaponization. Thus, the corrective logic of internalizing the shared cost of ruin and diluting the winner-takes-all characteristic of the game will mitigate the suicide region wherever this structure holds.

Ultimately, our findings map the AGI race onto the political economy framework of "The Narrow Corridor" (Acemoglu & Robinson, 2019). The path to safe deployment of AGI requires a "Red Queen" dynamic[16] where the advancement in AI safety research (society's shackles) must scale proportionally with technical capabilities. Currently, the incentive structure of the game ($L > F$) compels players towards early preemptive deployment of AGI without the corresponding safety mechanism in place ($V_P^* < V_t < V_S^*$), pushing the world out of the "narrow corridor" and toward the despotic outcome of an unshackled Leviathan – an AGI system unmoored from human control and prone to systemic misalignment. Re-entering the corridor requires not only technological progress, but mechanisms that are designed to require safety research as prerequisites for economic viability.

[16] The term refers to the "Red Queen" effect originally proposed in evolutionary biology by Van Valen (1973), derived from Lewis Carroll's Through the Looking-Glass: "Now, here, you see, it takes all the running you can do, to keep in the same place". Acemoglu and Robinson (2019) adapt this to political economy, arguing that liberty is not a static condition but a dynamic struggle where civil society's constraints must evolve at the same velocity as the state's coercive power.